\documentclass[twocolumn,aps,prb,superscriptaddress,floatfix]{revtex4-1}
\usepackage{graphicx}
\usepackage{amsmath}
\usepackage{tabularx}
\usepackage[caption=false]{subfig}
\begin{document}
\title{Breakdown of the Nagaoka phase at finite doping}

\author{Ilya Ivantsov}
\affiliation{Bogoliubov Laboratory of Theoretical Physics, Joint
Institute for Nuclear Research, Dubna, Russia}
\affiliation{L.V.Kyrensky Institute of Physics, Siberian Branch of Russian Academy of Sciences, Krasnoyarsk, Russia}
\author{Alvaro Ferraz}
\affiliation{International Institute of Physics - UFRN,
Department of Experimental and Theoretical Physics - UFRN, Natal, Brazil}
\author{ Evgenii Kochetov}
\affiliation{Bogoliubov Laboratory of Theoretical Physics, Joint
Institute for Nuclear Research, Dubna, Russia}

\begin{abstract}
The Nagaoka ($U=\infty$)
limit of the Hubbard model on a square lattice is mapped onto the itinerant-localized Kondo model
at infinitely strong coupling. Such a model is well suited to perform quantum Monte Carlo (QMC)
simulations to compute
spin correlation functions. For periodic boundary conditions, this model is shown to exhibit no short-range ferromagnetic (FM) spin correlations
at any doping $\delta\ge 0.01$ and at finite temperature, $T=0.1t.$ Our simulations give no indication that there is a tendency towards ferromagnetic ordering in the ground state, with more than one hole.
Employing on the other hand the open boundary conditions (or mixed boundary conditions) may result in the qualitatively different results for the thermodynamic limit depending on a way one chooses to approach  this limit.
These observations imply that the relevant thermodynamic limit remains unclear.

\end{abstract}
%\pacs x 74.20.Mn, 74.20.-z
\maketitle

\section{Introduction}

The strong electron correlations are at work to
full extent in the Nagaoka ($U=\infty$)
limit of the Hubbard model. Indeed, in this case an infinitely strong Coulomb repulsion strictly prohibits
the double electron occupancy of the lattice sites, and the no double occupancy (NDO) constraint becomes of the utmost importance.
In the infinite $U$ limit, the Hubbard Hamiltonian reduces to
\begin{equation}
H_{U=\infty}=-\sum_{ij,\sigma}t_{ij}\tilde c^{\dagger}_{i\sigma}\tilde c_{j\sigma},\quad \tilde c_{i\sigma}^{\dagger}
=c_{i\sigma}^{\dagger}(1-n_{i-\sigma}),
\label{1.1}\end{equation}
where $t_{ij}$ is a symmetric matrix whose elements represent the hopping amplitude $t>0$ between the
nearest-neighbour sites and which are, otherwise, zero.
Despite its seemingly simple form, this Hamiltonian cannot be diagonalized due to the fact that the
projected electron operators $\tilde c_{\sigma}$ fulfil  complicated commutation relations
resulting from the explicit
manifestation of strong correlations.

The physics behind the model (\ref{1.1}) is certainly far from  trivial.
Indeed, Nagaoka \cite{nagaoka} proved a theorem stating that for one hole
the ground state of the $U=\infty$ Hubbard model is a fully saturated
ferromagnet. This provides an interesting example of a quantum system
in which ferromagnetism appears as a purely kinetic energy effect with
hole hoping
(itinerant ferromagnetism)
emerging as a result of the strong correlations from the NDO constraint.

Unfortunately, despite extensive work over many years, this model and
itinerant ferromagnetism are still poorly understood.
One of the important questions  to be addressed concerns the
thermodynamic stability of the Nagaoka phase.
That is, whether or not the Nagaoka state is stable when
the density of holes is finite in the thermodynamic limit.
There are arguments both for \cite{RY89,LONG-rasetti-94,richmond,36,21,44,becca-sorella2001, mielke}
and against
\cite{suto,tian91,PLO92,tasaki}
the thermodynamic stability of the
Nagaoka phase
and comparisons between various approaches have been made carefully (for a
recent example, see \cite{kotl}).
The basic problem that prevents one from reaching a definite conclusion
on that
is  the large-$U$ limit or, equivalently,
the local NDO constraint  which
is very difficult to deal with in a controlled way.

Analytical approaches
basically imply a mean-field (MF) treatment
in which the local NDO constraint is uncontrollably
replaced by a global condition.
For example, the standard
slave fermion (SF) MF theory which treats the NDO
constraint only
on average is known to predict a
stable FM phase for the $U=\infty$ Hubbard model
over an unphysically large doping range.
It was however shown that the SF MF approach produces spurious results and it is therefore not reliable
for the description of the Nagaoka ferromagnetism.\cite{bfk}

Available
variational approaches
\cite{richmond,LONG-rasetti-94,shastry,vonderlinden-etal-91,basile-etal-90,hanisch-etal-97,becca-sorella2001}
show that
variational estimations
that involve
more realistic refined trial wave-functions result in a smaller
value of the critical hole concentration.
For example, by advanced analytical means, a rather small upper bound on the critical hole concentration, namely, $\delta_c=0.25$ was obtained. \cite{wurth} This result has been recently confirmed by the variational Monte Carlo investigation. \cite{carleo}
In case the mean-field treatment provides an exact result, i.e., in infinite spatial dimensions, the fully polarized FM ground state is never stable.\cite{fazekas}
One might therefore
think that a proper treatment
of the NDO constraint
could improve the MF results shifting the critical hole concentrations
towards progressively smaller values.

\section{Model}

In the present Section, we reformulate the standard infinite $U$- Hubbard model (\ref{1.1})  in a form that explicitly takes into account some basics facts concerning the physics of strongly correlated electrons at low doping. This enables us to apply numerical quantum Monte Carlo calculations in a more efficient way.

In the underdoped cuprates, one striking feature is the simultaneous
localized and itinerant nature of the lattice electrons.
One might hope therefore that representing the model (\ref{1.1}) in a form that takes
both aspects into consideration  on equal footing  would  help to address the problem in a more efficient way.
To this end,
Ribeiro and Wen proposed a slave-particle spin-dopon representation of the projected electron operators
in the enlarged Hilbert space \cite{wen},
\begin{eqnarray}
\tilde c_{i}^{\dagger}
=c_{i}^{\dagger}(1-n_{i-\sigma})=\frac{1}{\sqrt{2}}(\frac{1}{2}-2\vec S_i\vec\sigma)\tilde d_i.
\label{tildec}\end{eqnarray}
In this framework, the localized electron is represented by the lattice spin $\vec S\in su(2)$
whereas  the doped hole (dopon) is described by the projected hole operator,
$\tilde{d}_{i\alpha}=d_{i\sigma}(1-n^d_{i-\alpha})$.
Here $\tilde c^{\dagger}=(\tilde c^{\dagger}_{\uparrow},\tilde c^{\dagger}_{\downarrow})^{t}$ and
$\tilde d=(\tilde d_{\uparrow}, \tilde d_{\downarrow})^{t}$.

In terms of the projected electron operators, the constraint of no double occupancy (NDO) that encodes the essence of strong correlation takes on the form
\begin{equation}
\sum_{\alpha}(\tilde c_{i\alpha}^{\dagger}\tilde c_{i\alpha})+\tilde c_{i\alpha}\tilde c_{i\alpha}^{\dagger}=1.
\label{a} \end{equation}
It singles out the physical $3D$ on-site Hilbert space. Only under  this condition the projected electron operators are
isomorphic to the Hubbard operators.
Within the spin-dopon representation (\ref{tildec}), the NDO constraint reduces to a Kondo-type interaction,\cite{pfk}
\begin{equation}
\vec{S_i} \cdot
\vec{s_i}+\frac{3}{4}(\tilde{d}_{i\uparrow}^{\dagger}\tilde{d}_{i\uparrow}+
\tilde{d}_{i\downarrow}^{\dagger}\tilde{d}_{i\downarrow})=0,
\label{cnstr}\end{equation}
with $\vec
s_i=\sum_{\alpha,\beta}\tilde{d}_{i\alpha}^{\dagger}\vec\sigma_{\alpha\beta}\tilde{d}_{i\beta}
$  being the dopon spin operator.

The on-site operator $${\cal P}=1-\vec{S_i} \cdot
\vec{s_i}+\frac{3}{4}(\tilde{d}_{i\uparrow}^{\dagger}\tilde{d}_{i\uparrow}+
\tilde{d}_{i\downarrow}^{\dagger}\tilde{d}_{i\downarrow}), \quad {\cal P}^2={\cal P}$$ commutes with the Hamiltonian and projects
out the unphysical states. In terms of the projectors, Eqs. (\ref{tildec}) can be rewritten in the form,
\begin{equation}
\tilde c_{i\sigma}^{\dagger}=\sqrt{2}\, sign({\sigma})\,{\cal P}_i\tilde d_{i-\sigma}{\cal P}_i,
\label{stc}\end{equation}
where $sign(\sigma =\uparrow,\downarrow)=\pm.$
In view of this, we have
\begin{equation}
H_{U=\infty}={\cal P}\sum_{ij,\sigma}2t_{ij}\tilde d^{\dagger}_{i\sigma}\tilde d_{j\sigma}{\cal P},
\quad {\cal P}=\prod_i {\cal P}_i.
\label{1..1}\end{equation}
Equivalently, Eq.(\ref{1..1}) can be represented in the form of the
lattice Kondo model at dominantly strong  Kondo coupling, $\lambda \gg t$: \cite{pfk}
\begin{eqnarray}
H_{U=\infty}&=& \sum_{ij\sigma} 2t_{ij}
{d}_{i\sigma}^{\dagger} {d}_{j\sigma}
+ \lambda
\sum_i(\vec{S_i} \cdot
\vec{s_i}+\frac{3}{4}{n}^d_i),
\label{1...1}
\end{eqnarray}
where we have dropped the "tilde" sign of the dopon operators, as it becomes irrelevant in the presence of the NDO constraint.
In spite of the global character of the parameter $\lambda$, it  enforces the NDO constraint locally due to the fact that
the on-site physical Hilbert subspace corresponds to zero eigenvalues of the constraint, whereas the nonphysical subspace
is spanned by the eigenvectors with strictly positive eigenvalues.

The unphysical doubly occupied electron states are separated from the physical
sector by an energy gap $\sim\lambda$.
In the $\lambda\to +\infty$ limit, i.e. in the limit in which $\lambda$ is much larger than any other existing energy
scale in the problem, those states
are  automatically excluded from the Hilbert space.
Right at the point $\lambda=+\infty$ the proposed model (\ref{1...1}) is equivalent
to the $(U=\infty)$ Hubbard  model of strongly correlated electrons. Away from that point,
this model describes a phenomenological Kondo model in which
the strength of the correlation is controlled by $\lambda$.
In $1D$, Eq. (\ref{1...1}) reproduces
the well-known exact results for the $(U=\infty)$ Hubbard model.\cite{fk}

At this point a following remark is in order,
concerning the correspondence between the two limits
$U\to\infty$ and $\lambda\to\infty$. It might seem that this limit is equivalent to merely setting
$t_{ij}=0$ in Eq.(\ref{1...1}) which reduces the problem to the on-site one. This is however not true for the original Hubbard model, where the eigenfunctions  are strongly entangled and very complicated.

The point is that,
in the strong coupling limit ($\lambda/t\gg 1$), the local spin-spin correlator between conduction dopons and localized moments,
$\langle \vec S_i\cdot \vec s_i \rangle $, converges to a value of $-3/4\,\langle n_d\rangle$.
This is precisely cancelled out by the term $3/4\,n_d$ that enters the kinetic term in Eq.(\ref{1...1}).
Because of this the theory remains finite.
The energy per site takes on the form,
$$E=A+{\cal O}(\lambda ^{-1}).$$ It starts with a $\lambda$ independent term $A$ since all the unphysical states are excluded in this strong coupling limit. The leading term depends on $t$ in a nontrivial way, since in the large $\lambda$ limit the underlying Hilbert space is modified by the NDO constraint.
For example,
in $1D$, the leading term takes the form  \cite{fk}:
$$A=-\frac{2t}{\pi}\sin(\pi \delta),$$ where $\delta$ is a hole concentration. The corresponding sub-leading terms can be  found using the results for a strongly coupled $1D$ Kondo model. \cite{sigrist}

\section{Exact diagonalization: periodic boundary conditions}
\subsection{Large clusters with one and two holes}
\begin{figure}
\includegraphics[width=\linewidth]{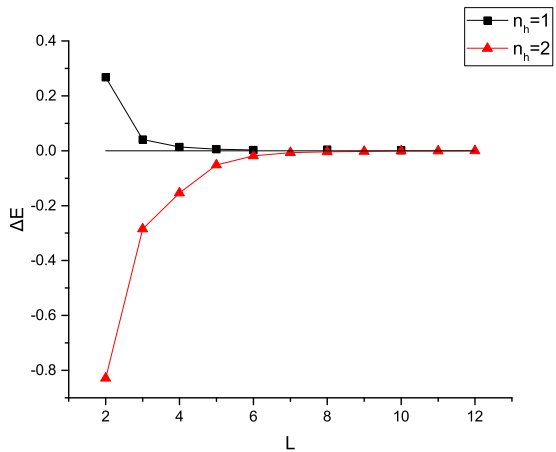}
\caption{The figure shows the energy difference $\Delta E=E(Q_{max}-1)-E(Q_{max})$ between the states with $Q=Q_{max}$ and $Q=Q_{max}-1$ for the different numbers of holes $n_h$. The lattice size $N=L\times L$ and the doping level $\delta=n_h/L^2$.}
\label{dE}
\end{figure}

\begin{figure*}
\begin{minipage}[h]{0.47\linewidth}
\center{\includegraphics[width=1\linewidth]{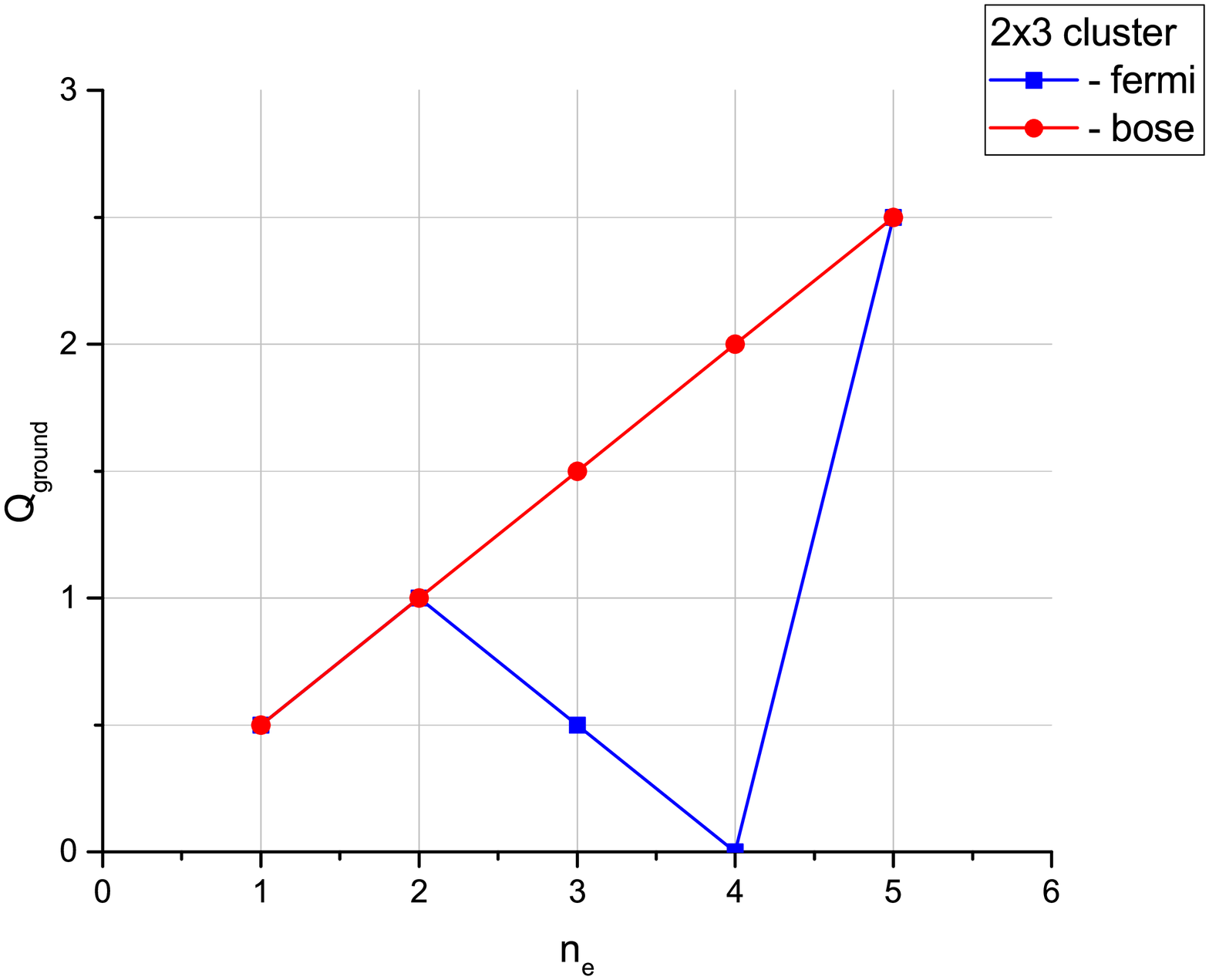}} a) \\
\end{minipage}
\hfill
\begin{minipage}[h]{0.47\linewidth}
\center{\includegraphics[width=1\linewidth]{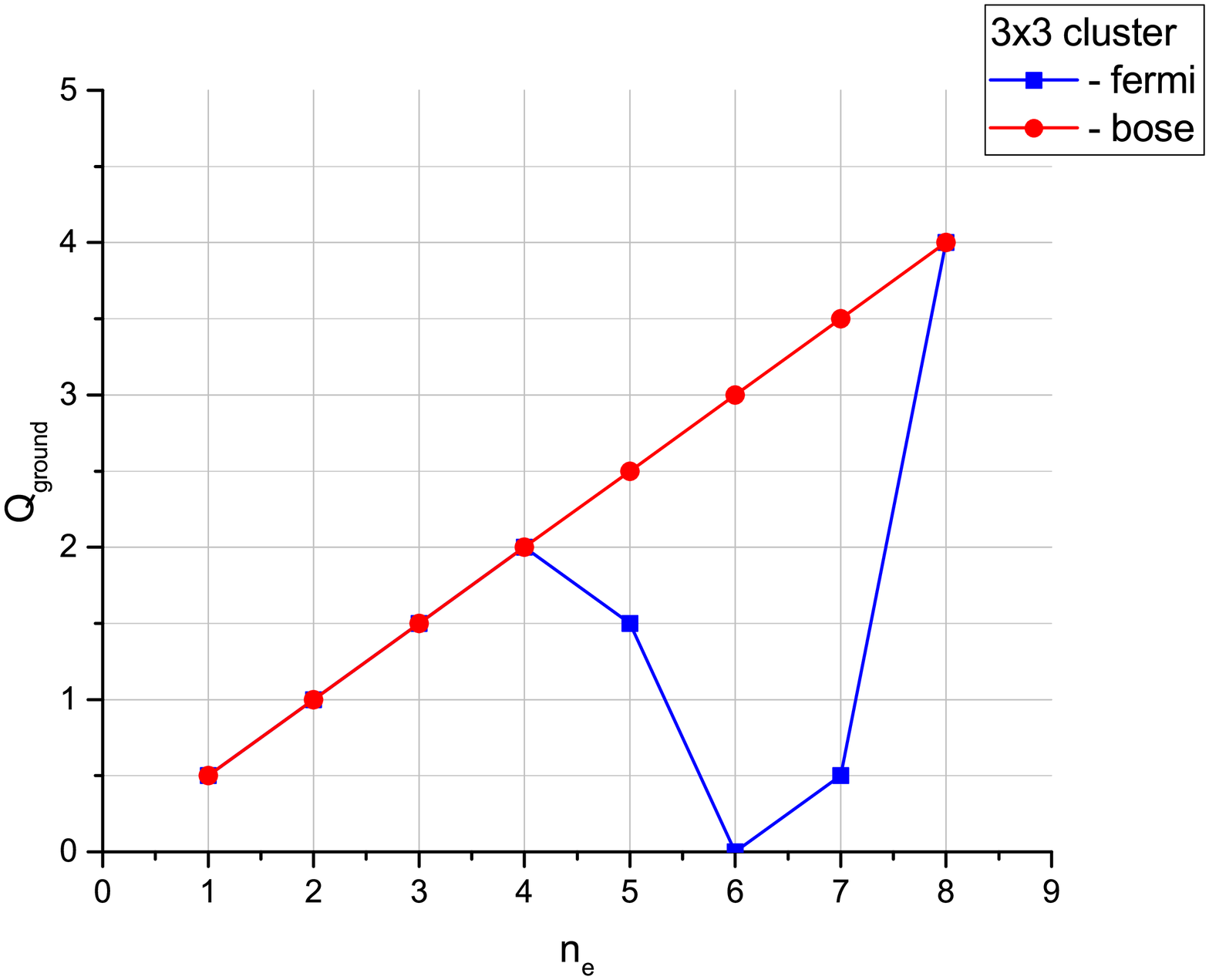}} \\b)
\end{minipage}
\vfill
\begin{minipage}[h]{0.47\linewidth}
\center{\includegraphics[width=1\linewidth]{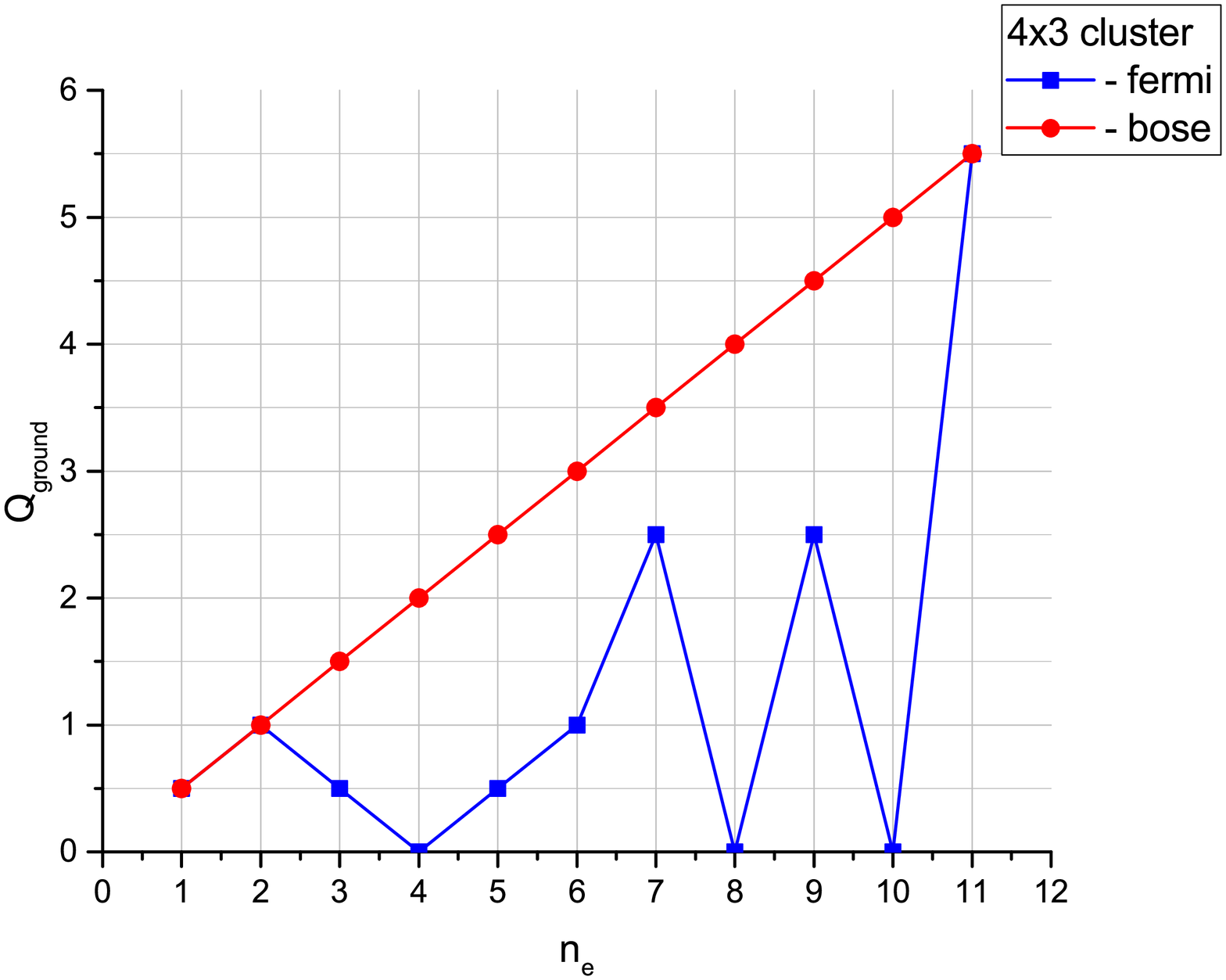}} c) \\
\end{minipage}
\hfill
\begin{minipage}[h]{0.47\linewidth}
\center{\includegraphics[width=1\linewidth]{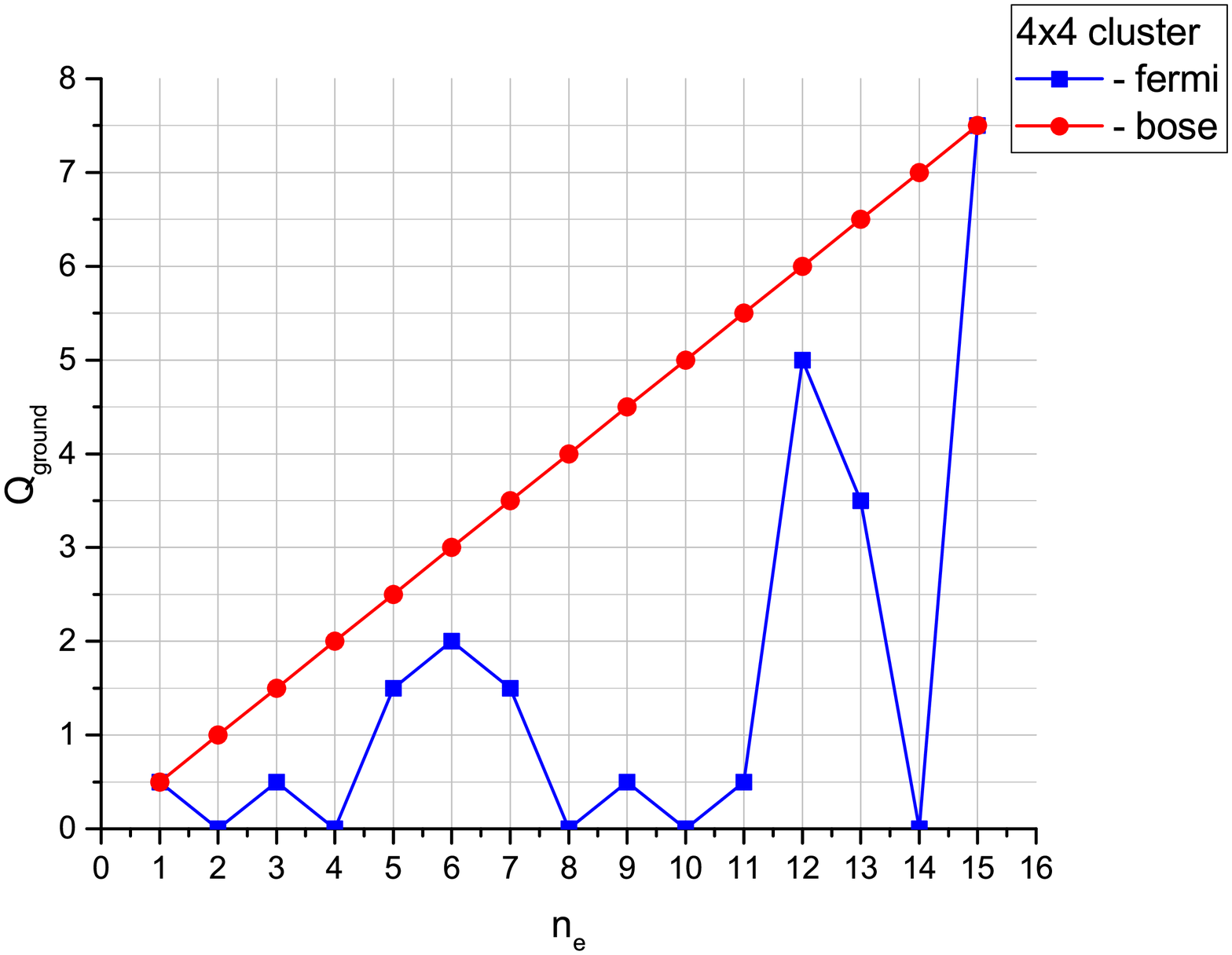}} d) \\
\end{minipage}
\caption{The spin value in the ground state for the different numbers of electrons $n_e=N-n_h$. Periodic boundary conditions are imposed.}
\label{Sg}
\end{figure*}

To get some insight into what kind of magnetic order one might expect at finite doping, we start by performing exact diagonalization of the finite lattice clusters. In this as well as in the two subsequent sections, we use  periodic boundary conditions (pbc). Some interesting and important modifications caused by the use of other  possible boundary conditions will be discussed in detail in Section V.

The size of the Hamiltonian matrix to be diagonalized  is $3^N\times3^N$, where $N$ is the number of sites.
This makes a diagonalization  difficult even for $N\sim20$.
However, the Hamiltonian is a block-diagonal matrix, where each block corresponds to a given number of electrons and a total spin projection.
Let us first restrict a number of holes to that of $n_h=1,2$ and fix the total spin projection to be $Q_z=Q_{max}, Q_{max}-1$.
In this case we can significantly enlarge the lattice size due to the fact, that the size of the largest Hamiltonian block ($n_h=2, Q=Q_{max}-1$) is proportional to the $N(N-1)(N-2)/2$.

In Fig.\ref{dE}, we report the results on the exact diagonalization of finite clusters with a maximal size up to $12\times12$ sites and the periodical boundary conditions (PBC).
The upper curve corresponds to the case of $n_h=1$ and it displays the difference between a fully polarized state and a state with one spin flipped, $\Delta E=E(Q_{max}-1)-E(Q_{max})$.
In case of one hole our results agree with Nagaoka's theorem, which predicts that a fully polarized state is the ground state.
The lower curve displays the same quantity for the case of $n_h=2$.
In case of two holes the energy of a fully polarized state lies higher than that of a state one spin flipped for all the considered lattices sizes.
Such a behaviour of the energy levels shows that the fully polarized state is not a ground state for two holes.
In other words, the Nagaoka state with two holes is unstable which fully agrees with results published elsewhere.\cite{Wen}

\begin{figure*}
\begin{minipage}[h]{0.47\linewidth}
\center{\includegraphics[width=1\linewidth]{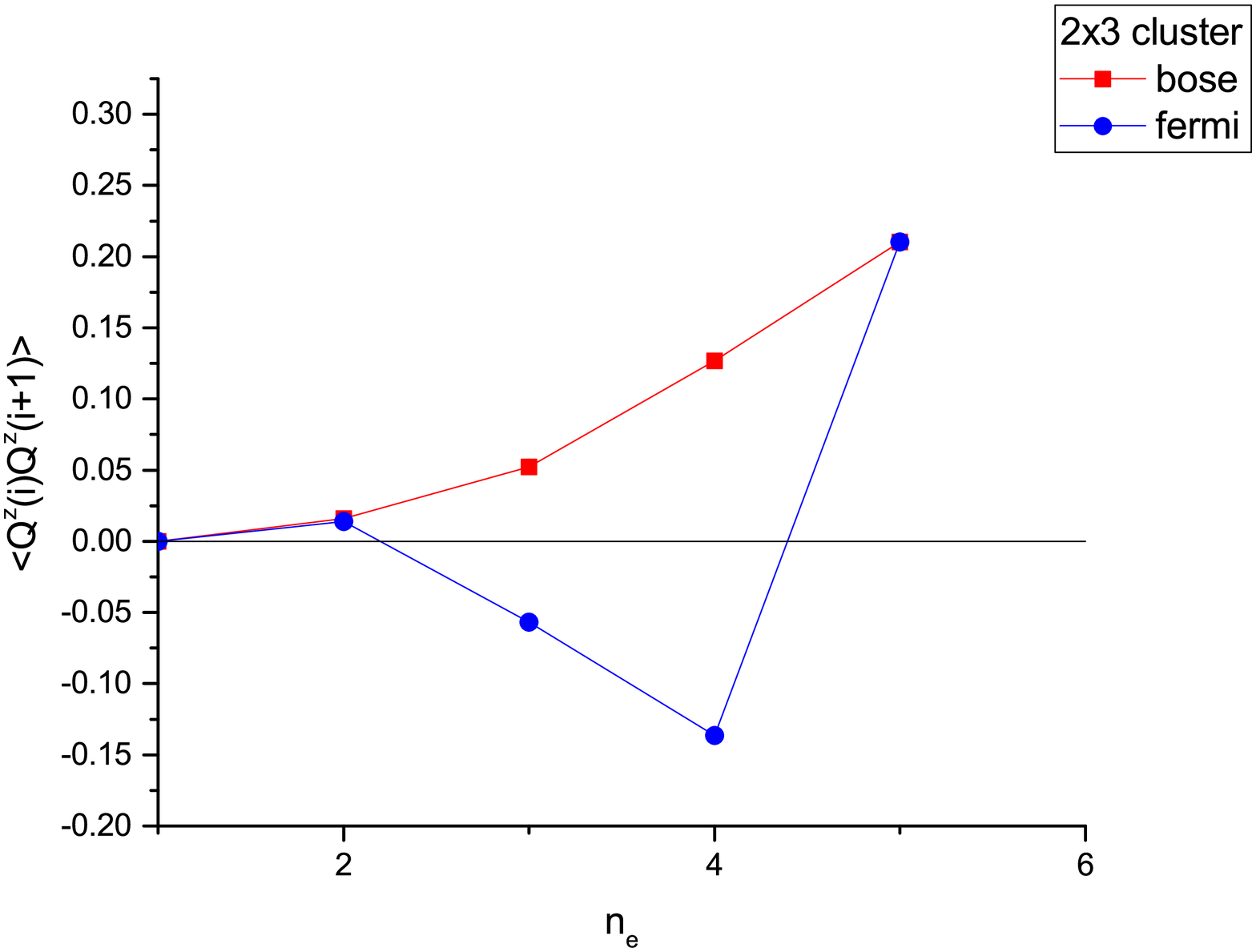}} a) \\
\end{minipage}
\hfill
\begin{minipage}[h]{0.47\linewidth}
\center{\includegraphics[width=1\linewidth]{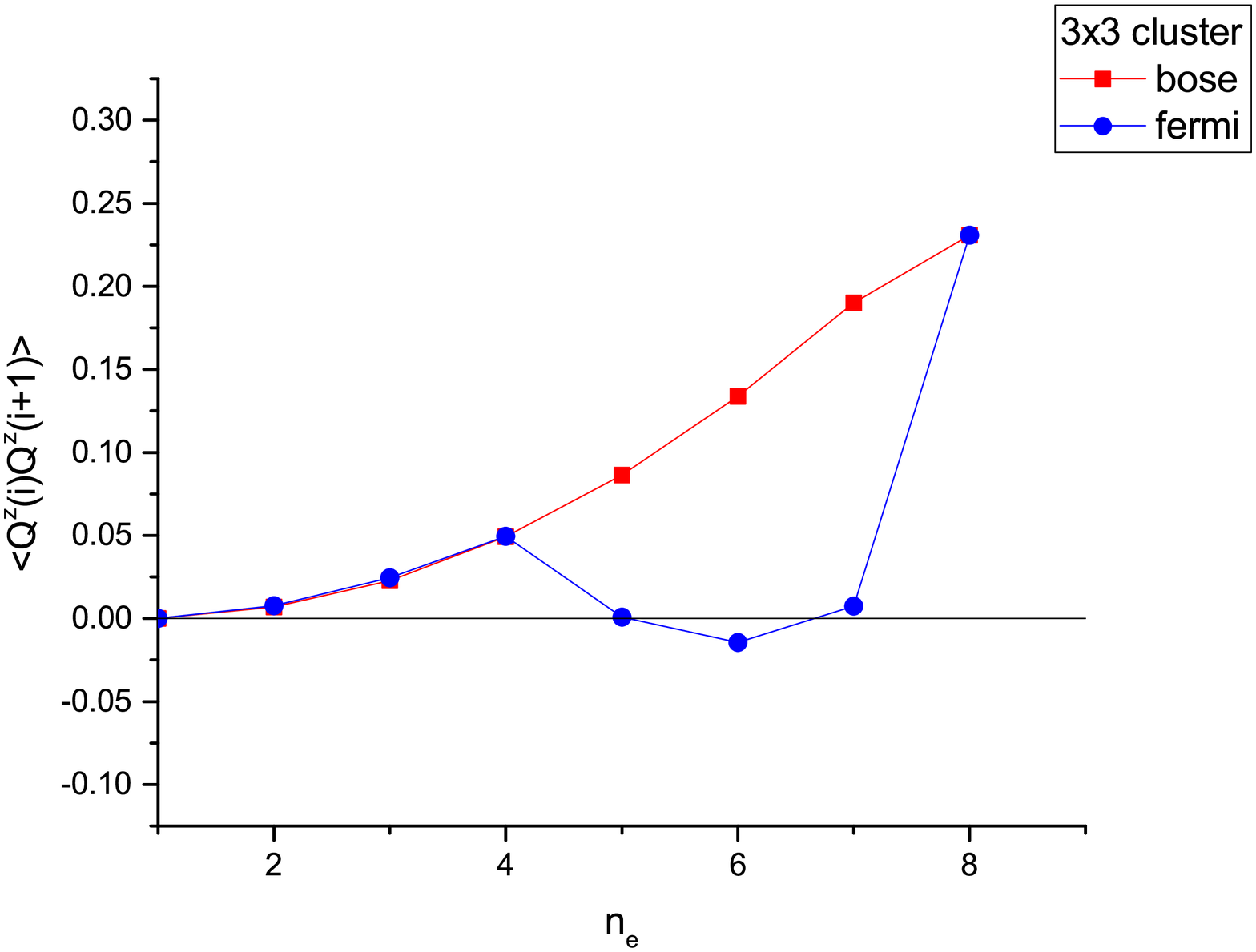}} \\b)
\end{minipage}
\vfill
\begin{minipage}[h]{0.47\linewidth}
\center{\includegraphics[width=1\linewidth]{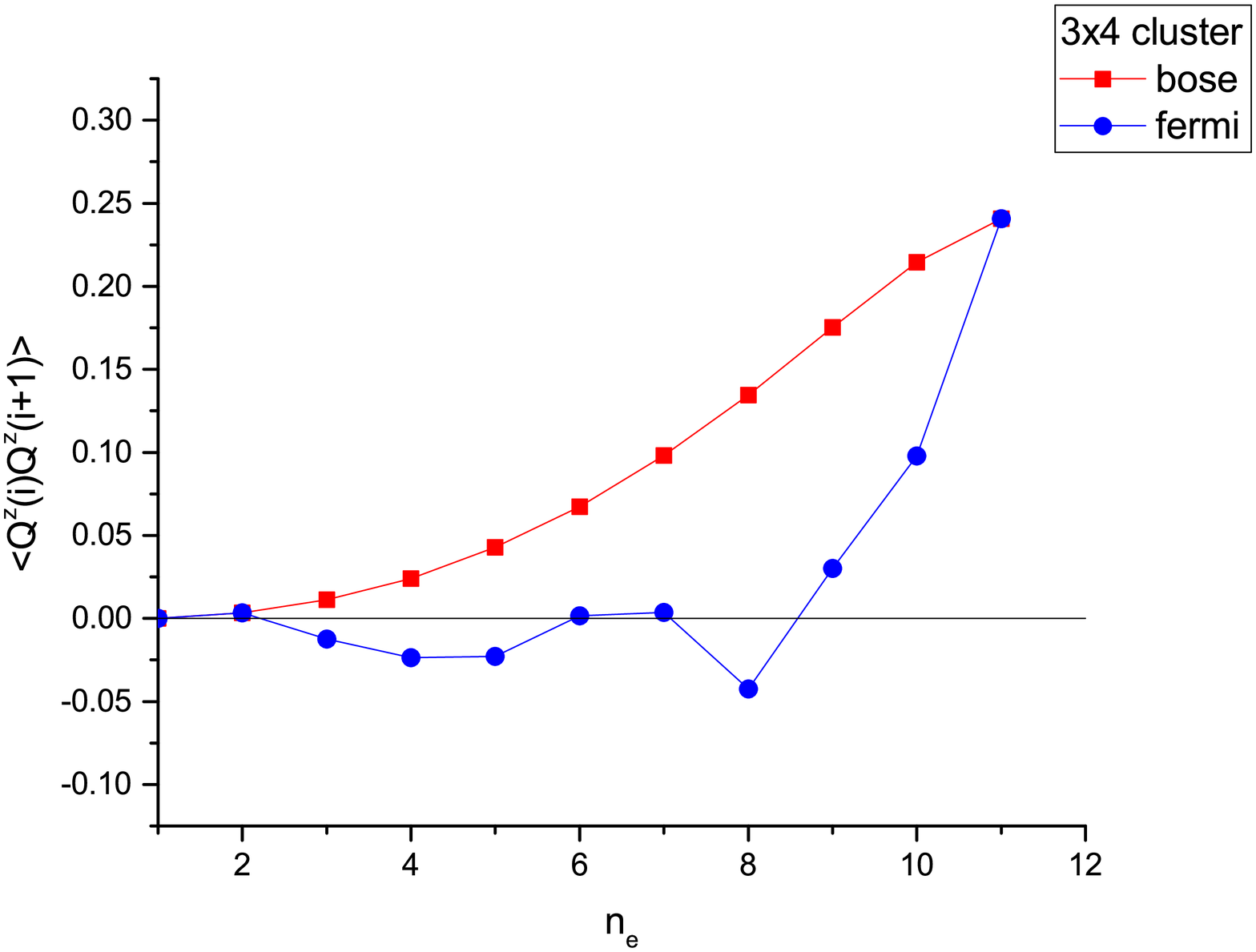}} c) \\
\end{minipage}
\hfill
\begin{minipage}[h]{0.47\linewidth}
\center{\includegraphics[width=1\linewidth]{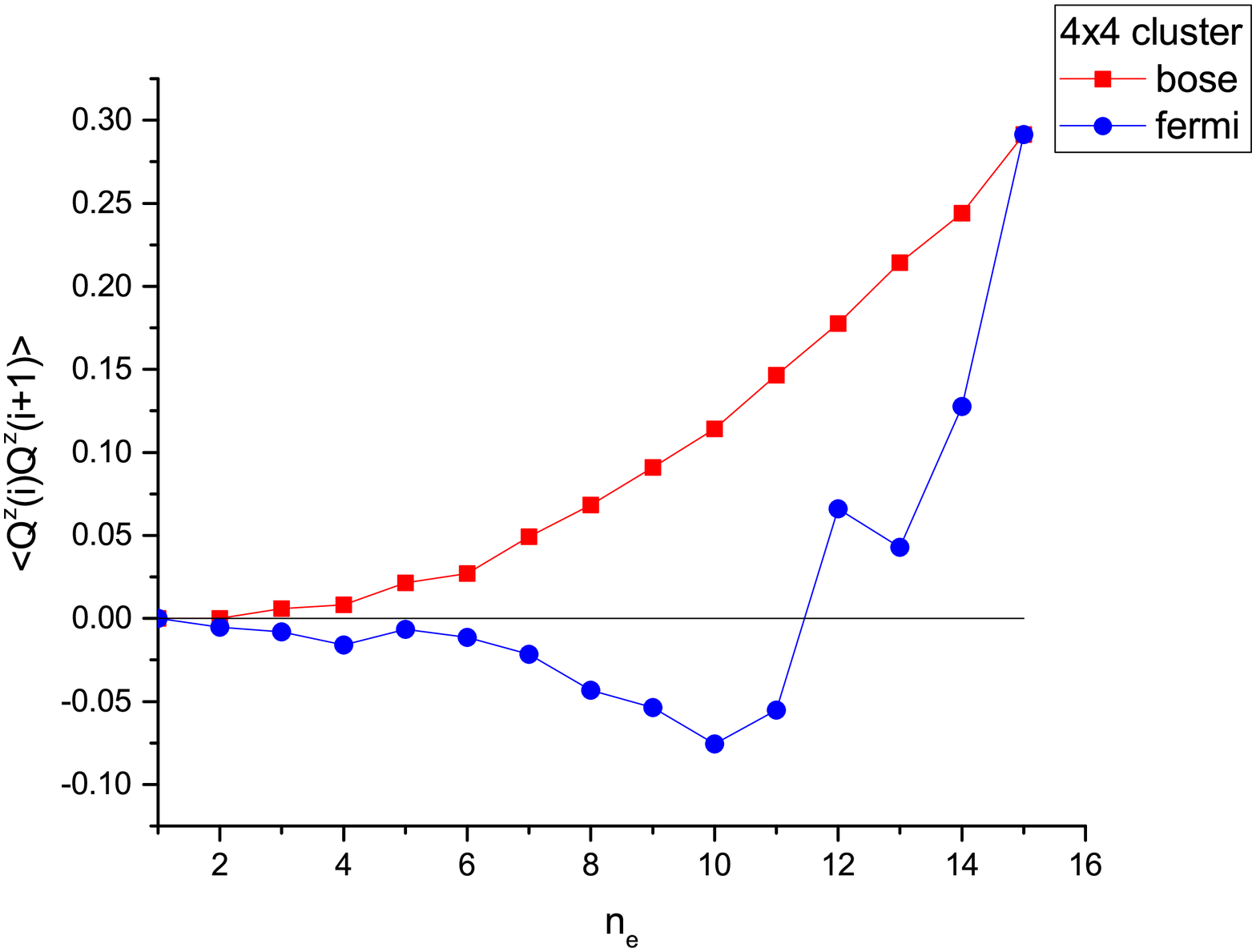}} d) \\
\end{minipage}
\caption{The $\langle Q^z_iQ^z_{i+1}\rangle$ correlation between nearest sites for the different numbers of electrons $n_e=N-n_h$. Periodic boundary conditions are imposed.}
\label{Gr}
\end{figure*}

\subsection{Small cluster exact diagonalization}

We further proceed by computing the lowest energy as a function of various hole numbers $n_h$ and total spin projections $Q_{ground}$ for different clusters
with the size up to $4\times4$. The obtained results presented in Fig.\ref{Sg} show that a fully polarized FM state is a ground state only if there is one hole in a lattice. For more than one hole, a fully polarized state is never a ground state.
Our results coincide exactly with those obtained by other methods. \cite{Takahashi}

On the other hand, for hard-core spinful bosons, the ground state appears as a fully polarized state at any number of holes.
This agrees \cite{muller} with the statement exactly proven elsewhere that, for spinful bosons, the
hard-core ferromagnetism  is stable for {\it all} hole densities.\cite{hcb}

Another interesting issue to address concerns the character of a possible associated quantum phase transition. Let us assume that there is a thermodynamically stable fully polarized state at finite doping in the region of $0\le \delta\le \delta_{cr}.$ Is then the onset of leading instability of this fully polarized phase
implies  that it occurs gradually, through small $\Delta Q=1$ changes in the total spin?  Or instead, such a transition
is discontinuous, meaning a large abrupt change in the total spin $\Delta Q\gg 1$? The results of the exact diagonalization of the small clusters
displayed in Fig.\ref{Sg} indicate that the transition from the Nagaoka state to te state  with two holes
always occurs through an abrupt spin change that increases with the lattice size. In particular, the results presented in the Panel (a) tell us that  $\Delta Q=2$, whereas those for the Panel (d) indicate that $\Delta Q=7.$ This therefore seems to indicate at first sight that
the breakdown of the Nagaoka phase at $T=0$ is  of a discontinuous character. However, the "oscillation" character of the curves displayed in Fig. 2 show that, at larger number of holes, $n_h>2$, there are in fact transitions with $\Delta Q=1$ or even $\Delta Q=0$.
It is not therefore clear what type of transitions actually survives in the thermodynamic limit.
In any event, the finite cluster calculations displayed in Fig.\ref{Sg} do not support the statement that the destruction of the Nagaoka state is necessarily
accompanied by an abrupt change of the total spin.

The exact small-cluster diagonalization can also be used to compute the spin-spin correlations $\langle Q^z_iQ^z_{i+1}\rangle$ between nearest-neighbour sites.
Figure \ref{Gr} shows these correlations  computed on the same clusters as those depicted in the Fig.2.
In the hard-core boson case, the correlation functions are purely ferromagnetic and scale as $n_e^2$.
This clearly corresponds to the fully polarized states.
For the hard-core fermions, however, the correlations do not reveal any FM magnetic order
except for the one-hole case, where the correlations coincide with those in the hard-core boson case.
In the fermionic case, a sort of AFM order builds up instead, with a magnitude of the correlations decreasing with
the total lattice size.
In the next section,
we confirm this result by QMC calculations for larger lattice clusters.

\section{Quantum Monte Carlo calculations}

An alternative approach is based on a computation of  spin-spin correlation functions by employing the QMC algorithm.
In this way, one can estimate a magnetic correlation length in the asymptotic regime, $r\gg a$.
In case this quantity shows no appreciable dependence on finite-size effects, one may draw certain reliable conclusions on a character of the underlying magnetic order.

One must however take a proper care of the fact that we are dealing
with a strong-coupling problem. Namely,
since we are interested in the large $\lambda$ limit, it seems appropriate to separate the Hamiltonian (\ref{1...1})
in the following way:
$H_{U=\infty}=H_0+H_{int},$ where the leading term is now
$$H_0=\lambda
\sum_i(\vec{S_i} \cdot
\vec{s_i}+\frac{3}{4}{n}^d_i),$$
and the "interaction" term is
$$H_{int}=2t\sum_{ij\sigma}
{d}_{i\sigma}^{\dagger} {d}_{j\sigma}, \quad t/\lambda\ll 1.$$

Although the exact diagonalization remains of the same complexity,
the spin-spin correlators can be computed
in a more efficient way by employing the localized and itinerant degrees of freedom displayed by our model
(\ref{1...1}). The convergence of the QMC  depends on a
selected basis.
Taking the Hamiltonian in this form allows us to achieve the significant weakening
of the sign problem in the low-doped case due to two facts.
First, it allows us to diagonalize the $H_0$ term in the one-site representation and
rewrite $H_{int}$ in the new basis.
During the simulation we can set $\lambda$  finite and then unphysical configurations that involve
$\lambda$-terms can occasionally occur. Since all measurements occur in the absence
of these configurations, they do not contribute to the final result.
This approach significantly improves the ergodicity of the algorithm and does not
contradict the detailed balance principle.
Secondly, in those circumstances the sign-problem weakens because of the fact that, in the spin-dopon representation, the
greater the density of dopons, the smaller is the average sign.
For example, in case of exactly one dopon, the sign-problem is absent. However if
standard electron operators are used the average sign in the same case is extremely small, and this is something we must definitely avoid.

To compute the spin correlators we adopt the continuous
time worldline (CTWL) QMC algorithm based on the representation of the partition function in the interaction picture:
\begin{equation}
e^{-\beta H_{U=\infty}} =e^{-\beta H_0} T_{\tau}(exp(-\int_0^\beta H_{int}(\tau)d\tau)),
\label{QMC}
\end{equation}
where $T_\tau$ stands for the $\tau$-ordering operator, and the partition function $Z_{U=\infty}=tr e^{-\beta H_{U=\infty}}.$
All the necessary details for the  CTWL QMC algorithm to be applied
to treat the lattice Kondo-Heisenberg model can be found in a recent paper  published elsewhere. \cite{ifk}
The numerical simulations exposed in that work are simply to be restricted to the case  of
$J=0$.

\begin{figure}
\begin{minipage}[h]{\linewidth}
\center{\includegraphics[width=1\linewidth]{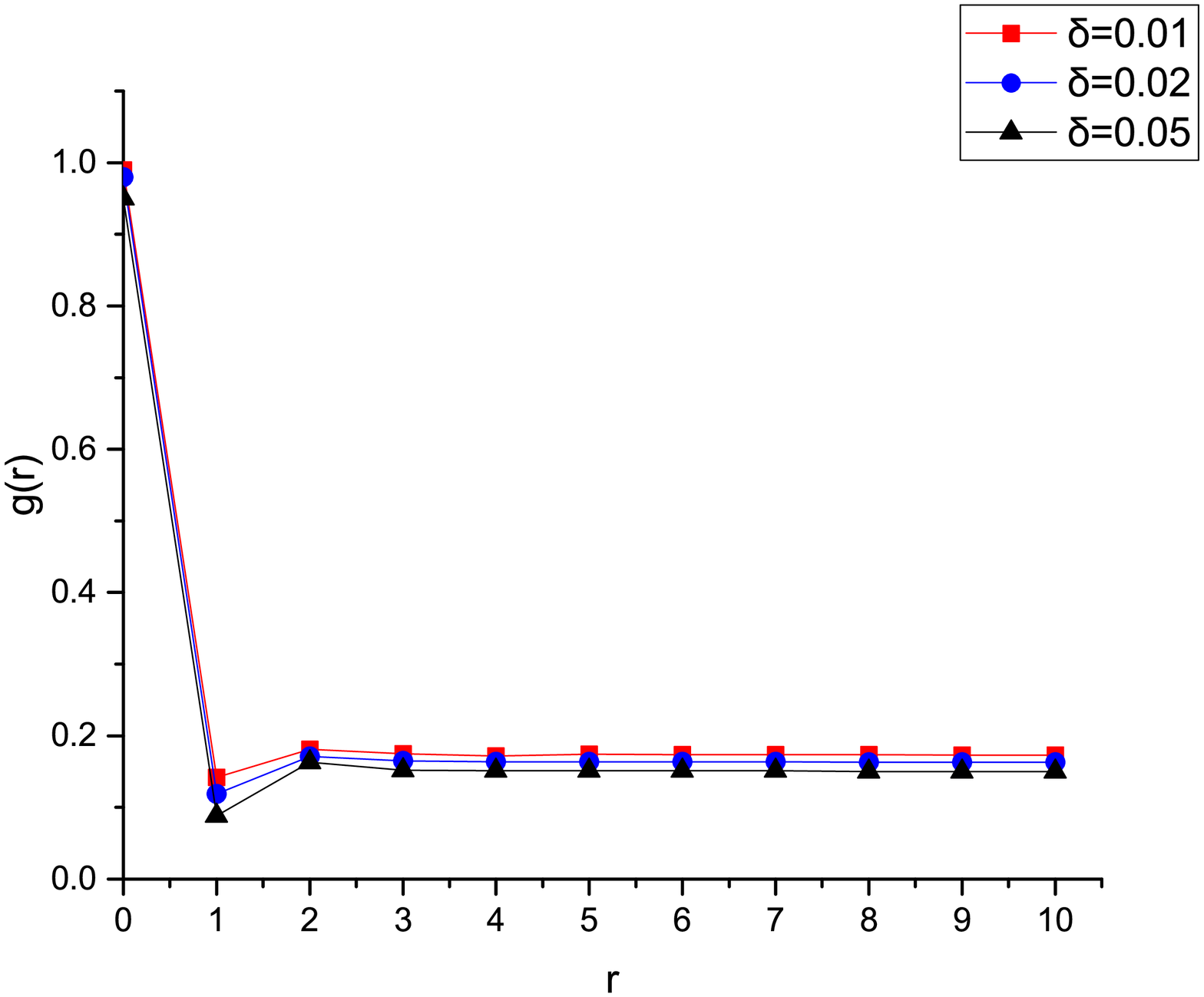}} a) \\
\end{minipage}
\vfill
\begin{minipage}[h]{1\linewidth}
\center{\includegraphics[width=1\linewidth]{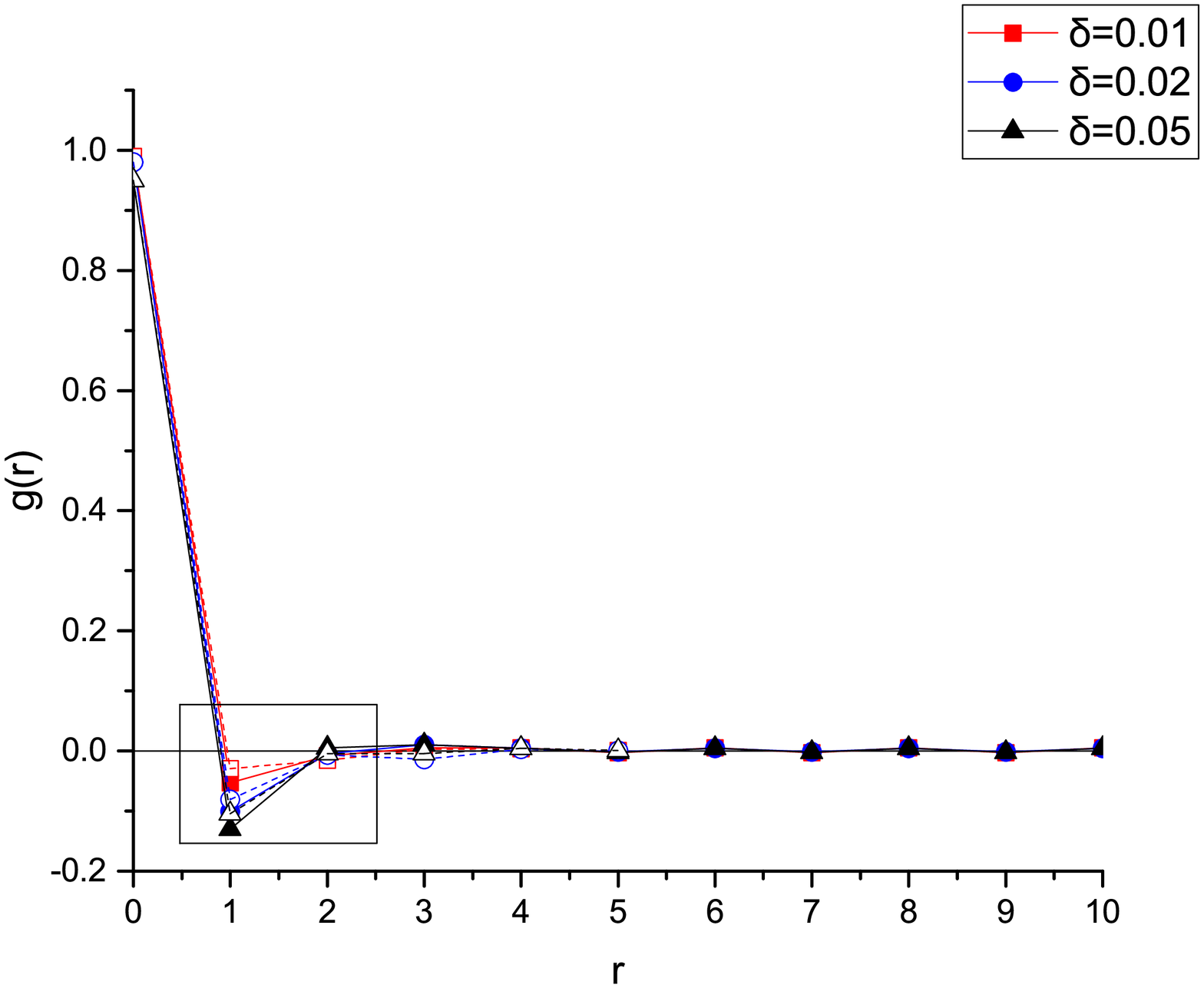}} \\b)
\end{minipage}
\vfill
\begin{minipage}[h]{1\linewidth}
\center{\includegraphics[width=1\linewidth]{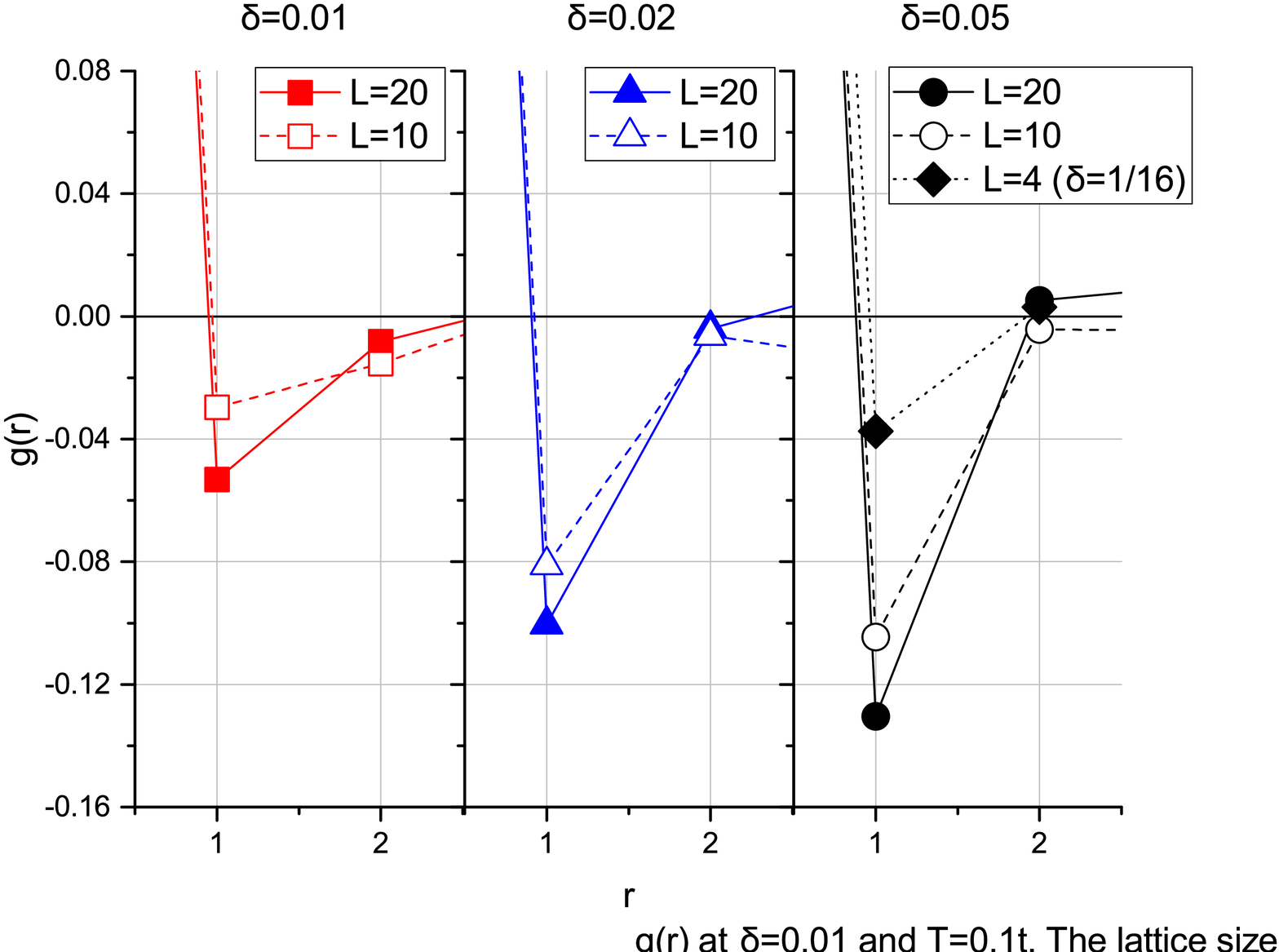}} c) \\
\end{minipage}
\caption{Panel (a) shows correlation functions for $L=10$ for different doping levels for the hard-core boson case at $T=0.1t$. Panel (b) shows the same for fermion case. Solid (dashed) lines show results obtained for $L=20$ ($L=10$), respectively. Number of lattice size is $N=L\times L$. Panel (c) shows the highlighted fragments of panel (b) with the doping level being separately presented .}
\label{gr}
\end{figure}

The  spin-spin correlation function $g(r)$
for the physical electron operators
is calculated for  a fixed number of dopons, $\delta$:
\begin{eqnarray}
\begin{gathered}
g(r)= \frac{4}{\Delta(r)}\sum_{ij}\langle (S^z_i+s^z_i) (S^z_j+s^z_j)\rangle \bar{\delta}(r-|\mathbf{R}_i-\mathbf{R}_j|),
\label{1.3}
\end{gathered}
\end{eqnarray}
where  $\mathbf{R}_i$ is the radius-vector of the site $i$, $\Delta(r)=\sum_{ij} \bar{\delta}(r-|\mathbf{R}_i-\mathbf{R}_j|)$ and
\begin{equation}
\bar{\delta}(x)=
\begin{cases}
 1, &\text{if $|x|\le 0.5a$,}\\
 0, &\text{otherwise,}
\end{cases}
\end{equation}
with $a$ being the lattice constant and $\langle...\rangle$ being an average over the spin configurations generated in the QMC run.
The corelation function is normalized by the condition $g(r=0;\delta=0)=1.$
All numerical results are obtained for a $20\times20$ lattice cluster with periodic boundary conditions.

Figure \ref{gr} displays the corresponding spin-spin correlators for the hard-core bosons (Panel (a))
and the hard-core fermions (Panels (b) and (c)).
In the boson case, the FM order is clearly observed in full accordance with the exact result \cite{muller}.
The small-cluster exact diagonalization results displayed in Fig.3. clearly show that
the correlation function scales at $T=0$ as $n_e^2$. This implies that $g(r)\approx 1$  at $\delta\ll 1$.
Finite-temperature effects considerably suppress the correlation function
as seen in Fig.\ref{gr}. However, it remains finite, since $g(r)\approx 0.2, \, r\gg a.$
The observed order in hard-core bosons case
is of course not truly long-ranged since it does not exist in $2D$ at finite temperature
but rather quasi long-range feature. The associated FM correlation length is finite but much larger than the cluster size. In fact it scales exponentially with $1/T$ as $ T\ll t$.

Panel (b) displays the fermion spin correlators for the different doping levels.
In sharp contrast to the hard-boson case, there is no tendency towards FM ordering at finite doping.
The spin-spin correlations in real space show no evidence of even short-range FM correlations,
but rather weak AF correlations instead.\cite{hirsch}
In case there indeed were a continuous phase transition at a critical doping $\delta_c$ at which a true long-range FM order in a ground state in $2D$
does emerge, then, in the associated  quantum critical region specified in particular by the requirements $\delta=\delta_c, \, T<< t$, the system must necessarily
display a finite  FM spin-spin correlation length that scales as $T^{-1/z}$ where $z\ge 1$ is a dynamic exponent.
This exponent parametrizes the relative scaling of space and time.
The precise value of $z$ could have been guessed provided an appropriate effective action to describe the low-energy
physics of the $U=\infty$ Hubbard model would have been available, which however is not the case.
Just to get an idea as to what might be the order of magnitude of the spin-spin correlation length away from the critical point,
let us assume that $z=1$ (In Fermi-liquid-like systems this would imply that there are no overdamped modes associated
with an ordering field \cite{hertz}).
In this case $\xi_{FM} \propto \lambda_{T},$
where the de Broglie wavelength at finite temperature $\lambda_{T}=v/T.$
This relation implies that thermal and quantum fluctuations are equally important in this (critical) regime.
The characteristic velocity of the quasiparticle low-energy excitations $v\propto ta$.
At a given temperature, $T=0.1t$, we therefore get
$\xi_{FM} \propto 10a$. In other words, the long-range magnetic order in the ground state would manifest itself at finite temperature
through a finite correlation length of  order of at least a few lattice spacings.
We however never observe  the spin-spin correlation functions
that display such a behaviour down to a very small doping level of $\delta=0.01$.

\section{Discussion: Role of boundary conditions }

\begin{figure*}
\begin{minipage}[h]{0.32\linewidth}
\center{\includegraphics[width=1\linewidth]{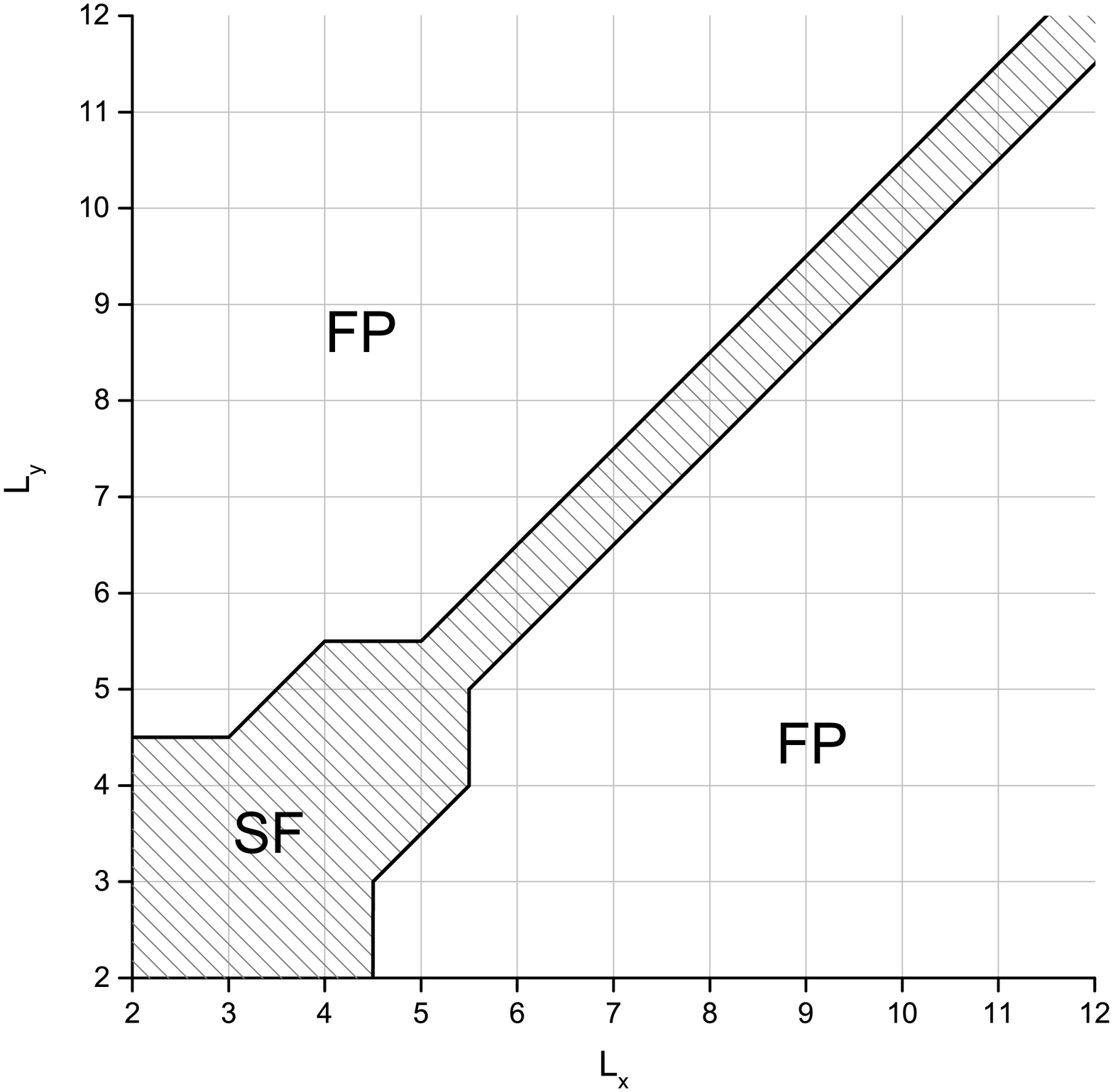}} a) \\
\end{minipage}
\hfill
\begin{minipage}[h]{0.32\linewidth}
\center{\includegraphics[width=1\linewidth]{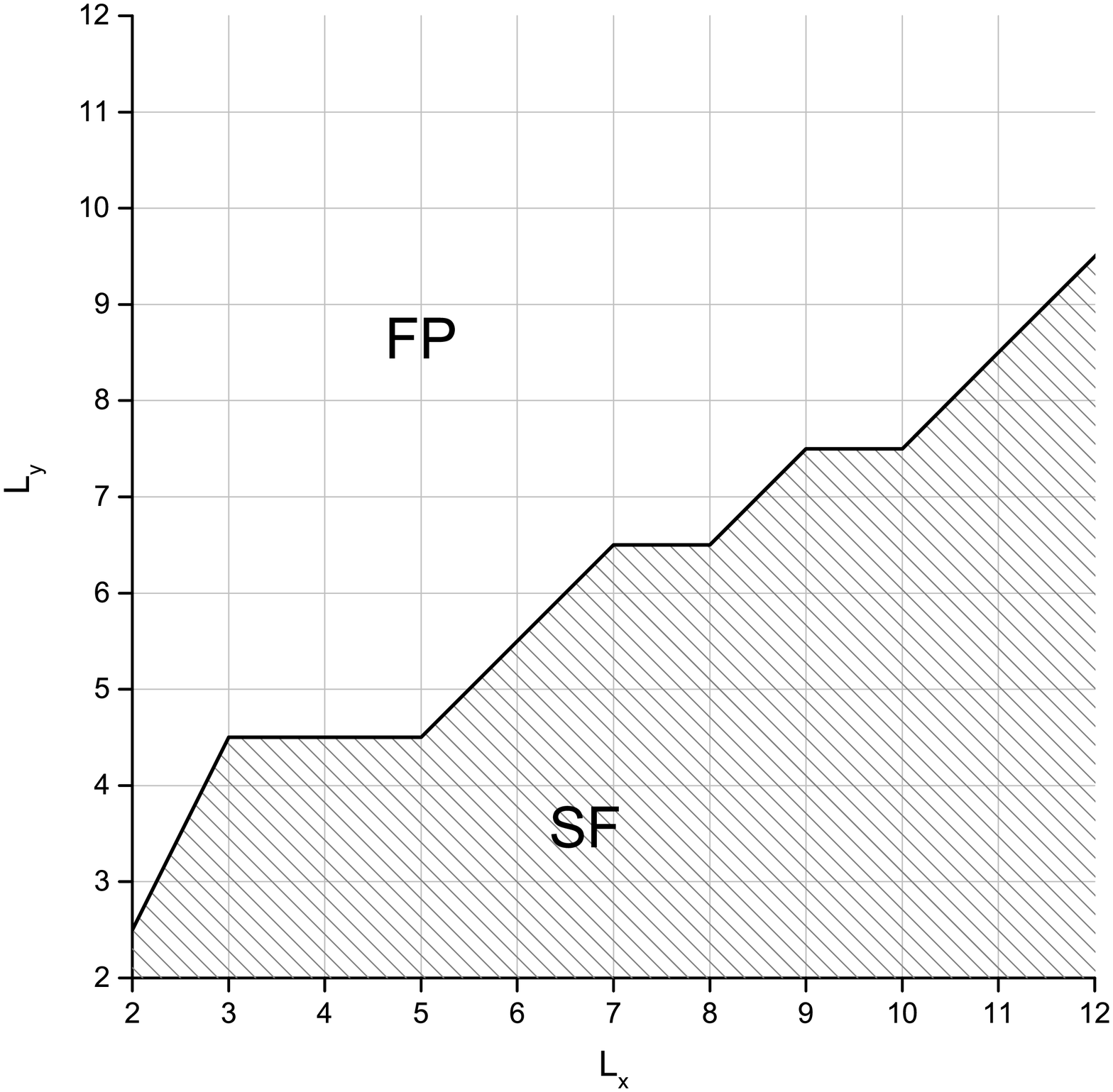}} \\b)
\end{minipage}
\hfill
\begin{minipage}[h]{0.32\linewidth}
\center{\includegraphics[width=1\linewidth]{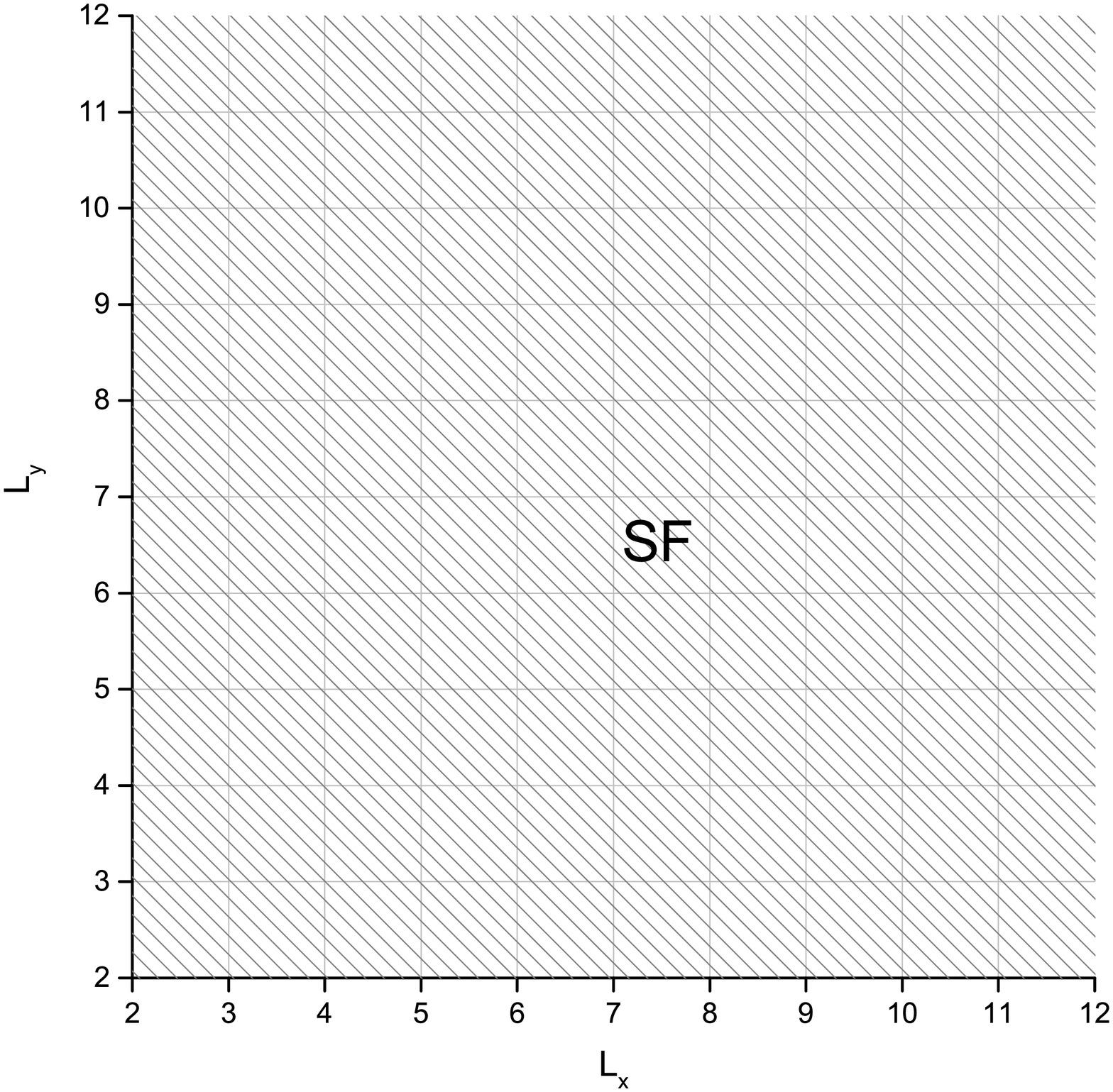}} \\c)
\end{minipage}
\caption{The figure shows the energy difference between the fully polarized (FP) state ($Q=Q_{max}$) and the one spin flipped (SF) state ($Q=Q_{max}-1)$) in case of two holes for the different lattice sizes $N=L_x\times L_y$. The SF areas correspond to the case $E(Q_{max}-1)<E(Q_{max})$, the FP areas correspond to the case $E(Q_{max}-1)>E(Q_{max})$. Panel (a) shows the lattice with the open boundary conditions, panel (b) shows the lattice with the periodic boundary condition for the axis $x$ and the open boundary condition for the axis $y$, panel (a) shows the lattice with the periodic boundary conditions}
\label{Phase}
\end{figure*}

We have shown that the spin-spin correlation function analysis provides strong arguments against thermodynamic stability of the Nagaoka phase. However, a different conclusion was reached in
a recent work \cite{white} that employs a density matrix renormalization group (DMRG)
approach to study a phase diagram of the infinite $U$ Hubbard model on $2$- to $6$- leg ladders.
The authors find a fully polarized FM phase at zero temperature, when $\delta$, the density of holes per site, is in the range $0< \delta <0.2$. As those results are largely insensitive of the ladder width, they consider them representative of the $2D$ square lattice.
These two conclusions  seemingly contradict each other.
In this Section, we intend to clarify this issue.

The important distinction between our work and the paper \cite{white} has to do with the choice
of boundary conditions. In the latter open boundary conditions (OBCs) are used whereas we use the PBCs instead.
We check by the exact diagonalization on rather large clusters how significant is this difference. To this end, we consider $12\times 12$ clusters with one and two holes. As we showed in Section III, in this case, the PBCs imply that the two-hole Nagaoka (fully polarized) state is never a ground state of the system.
In case of the OPB a one hole state is again fully polarized as it should be as is in the case  for the PBC in agreement with Nagaoka's theorem.
However, the two-hole case  is now represented by the two distinct phases.

Fig.\ref{Phase} shows the phase separation for the different lattice sizes and different boundary conditions, with
the panel a) corresponding to the OBCs.
The fully polarized (FP) phase implies that the $E(Q=Q_{max})<E(Q=Q_{max}-1)$.
In this case, the FP ground state is allowed though it is not necessarily realized.
In contrast, the one spin flipped (SF) phase depicted by the shaded area corresponds to the case of $E(Q=Q_{max})>E(Q=Q_{max}-1)$.
This phase strictly prohibits a fully polarized ground state, because an energy of at least one spin flipped state is lower than that of the fully polarized state.
The phase diagram in Fig. 5a) is in an accordance with the results of Ref.\cite{white}
At sufficiently large length of the $2,4,6$-leg chains  a fully polarized $2$-hole state is  lower than
that with one spin flipped.

Our calculations agree with those obtained within the DMRG approach for large enough chain length.  For example, at the critical hole concentration $\delta_c=0.2$, a two-leg chain exhibits a fully polarized ground state only when the chain length is equal to or larger than $5$. In this case, the critical number of the doped holes
at which the FP state becomes more favourable should be equal to
$(2\times 5)\times (1-0.8)=2$, which agrees with the results depicted in 5a).

We thus see that the OBCs allow for the existence of the fully polarized ground state at a sufficiently large lattice ( a length along one of the axes $\ge 5$). There is however one exception. As seen from Fig.5a), the square $L\times L$ lattice clusters with two holes have at least one state, which has lower energy that the fully polarized one. The Nagaoka state is unstable in this case. This is an interesting observation, since usually just square lattice clusters are used in the actual carrying out the $2D$- limiting procedure.

Let us now turn to the case of the so-called mixed boundary conditions (MBCs). The results are reported in Fig.\ref{Phase}b). The MBC imply that the PBCs are imposed along the $x$- axis and the anti PBCs -- along the  $y$-axis. As seen from the panel b), in case the periodicity holds along the "long" side of a lattice rectangle, than the SF state is more favourable; in the opposite case  -- the fully polarized state
is preferable. This is quite natural, since if a $2\times 50$ lattice lattice cluster is bent into a ring along the short side so that only two sites on the opposite sides interact, this clearly will produce  no noticeable effects.

Finally, Fig.5 c) shows
that the PBC used in our work imply that the state with $Q=Q_{max}-1$ is always more favourable  than the  polarized one. In other words, the fully polarized state with the  PBCs is never a ground state.

\section{Conclusion}

In conclusion, we show that there is an influence of the boundary conditions on the thermodynamic limit for the $(U=\infty)$ Hubbard model.
The contribution that comes from the fixing boundary conditions may play a key role even for very large lattice cluster calculations.
In principle, this happens because such contribution and that which comes from the difference in energy between a fully polarized and the unpolarized states are of the same order -- ${\cal O}(1/N)$ \cite{Wen}.

Within the approach based on the PBCs (for which the original Nagaoka theorem was formulated),
the lack of a clear sign of FM short-range spin correlations in the parameter range studied in our work
provides strong arguments against thermodynamic stability of the Nagaoka phase at least
for  the hole concentrations $\delta\ge 0.01$.
It is very likely to conclude that the critical hole concentration
is in fact equal to zero.
There are actually  only
two acceptable options in the thermodynamic limit in this case, namely,  either the FM order does not exist at all,
which is the case for the hard-core fermions, or it
is realized at all possible hole densities.
This last (unphysical) option corresponds to implementing the hard-core bosonic statistics for the constituent particles.
Employing on the other hand the OBCs (or MBCs) may result in the qualitatively different results for the thermodynamic limit depending on a way one chooses to approach   this limit
(by using square or rectangular building blocks, e.g., in case of the OBCs).
These observations imply
that the relevant thermodynamic limit remains unclear.

\section{Appendix}

In the Appendix, we provide a derivation of the Nagaoka theorem within a framework of the spin-dopon representation.
For simplicity, we
restrict ourselves to the case of a $D$-dimensional hypercubic regular lattice. In this case the sign of $t$ is
irrelevant so that we can fix the Hamiltonian to be
\begin{eqnarray}
H_{U=\infty}
&=&-2t\sum_{ij\sigma}{d}_{i\sigma}^{\dagger} {d}_{j\sigma}\nonumber\\
&+&\frac{3\lambda}{4}\sum_{i\sigma}{d}_{i\sigma}^{\dagger} {d}_{i\sigma} +\lambda
\sum_{i}\vec{S_i} \cdot \vec{s_i},\quad t> 0.
\label{7}\end{eqnarray}
The limit $\lambda\to \infty$
reduces the on-site Hilbert space to that comprising  a spin-up state $|\uparrow \rangle_i=|\uparrow 0\rangle_i$,
a spin-down state
$|\uparrow \rangle_i =|\uparrow 0\rangle_i$ and  a vacancy state $|0\rangle_i=\frac{|\uparrow\downarrow\rangle_i-
|\downarrow\uparrow\rangle_i}{\sqrt{2}}$. We should therefore consider
\begin{eqnarray}
H=-2t\sum_{ij\sigma}{d}_{i\sigma}^{\dagger} {d}_{j\sigma}
\label{8}\end{eqnarray}
in the reduced Hilbert space.

We define the basis one-vacancy states as
$$|i,\{\sigma\}\rangle=|\sigma_1\sigma_2...0_i...\sigma_N\rangle,$$
where $\sigma_k=\uparrow\downarrow$ and $\{\sigma\}$ is a multi-index describing an arbitrary set
of the lattice spins.
The vacancy state $|0\rangle_i$ is a total spin singlet defined above.

Let
$$|\Psi\rangle=\sum_{(i,\{\sigma\})}\psi_{(i,\sigma)}|i,\{\sigma\}\rangle$$ be an arbitrary one-hole normalized state.
Let us define a state
$|\Phi\rangle$ with $Q=Q_{max}=Q^z=(N-1)/2$ as
\begin{equation}
|\Phi\rangle=\sum_i\phi_i|i,\{\uparrow\}\rangle.
\label{9}\end{equation}
Here $\vec Q=\sum_i(\vec S_i+\vec s_i)$ is a vector of the total spin of the electrons,
$\phi_i=(\sum_{\sigma}|\psi_{i,\sigma}|^2)^{1/2}$, and the multi-index
$\{\uparrow\}$ represents that all the lattice spins point upwards.
The energy of the state $|\Psi\rangle$ is evaluated to be
\begin{eqnarray}
\langle\Psi|H|\Psi\rangle&=&-t\sum_{ij,\sigma\tau}\bar\psi_{i\tau}\psi_{j,\sigma}\ge
-t\sum_{ij}\bar\phi_j\phi_i\nonumber\\
&\ge&-t\sum_{ij}|\phi_i|^2=-tz,
\label{10}\end{eqnarray}
where $z$ is a coordination number. To obtain (\ref{10}) we have repeatedly used the Schwartz inequality. \cite{tasaki1}
From (\ref{9}) it follows that the energy of the ground state
$$E_{gr}=-tz.$$
The last inequality in (\ref{10}) is saturated for the state (\ref{9}) provided $\phi_i=const=1/\sqrt{N}$.
Such a state describes the fully polarized lattice spins
and a hole with the highest mobility.
This indicates that the fully polarized ferromagnetic state
$$|\Phi_{gr}\rangle=\frac{1}{\sqrt{N}}\sum_i|i,\{\uparrow\}\rangle$$ is indeed the ground state of the system.
This state has $Q=Q^z=(N-1)/2$.

Now it is left to show that there is no another state with $E=-tz$ and $Q < Q_{max}$.
Let us denote the state with $E=-tz$ and arbitrary given $N_{\uparrow}$ and $N_{\downarrow}$ by
$$ |\Phi\rangle =\sum_{(i\{\sigma\})}\psi(i\sigma)|i,\{\sigma\}\rangle.$$ The Schr\"odinger equation
$$H|\Phi\rangle=-tz|\Phi\rangle$$ then gives
\begin{equation}
\psi(i\sigma)=z^{-1}\sum_{j\tau=n[i\sigma]}\psi(j\tau).
\label{11}\end{equation}
Here $n[i\sigma]$ denotes nearest neighbours of $(i\sigma)$.
The unique solution to (\ref{11}) reads \cite{nagaoka}
\begin{equation}
\psi(i\sigma)=const.
\label{12}\end{equation}
This again corresponds to the state with $Q=Q_{max}$ and $Q_z=(N-1)/2$, so that
there is no state with $E=-tz$ and $Q < Q_{max}$.


\begin{thebibliography}{99}
\bibitem{hcb} The hard-core boson model is given by Eq.(\ref{1.1}) in which the hard-core fermion operators $\tilde c_{i\sigma}$
are replaced with the hard-core boson ones, $b_{i\sigma},\, \sum_{\sigma}b^{\dagger}_{i\sigma}b_{i\sigma}\le 1.$
\bibitem{hirsch} Within the QMC simulations of the
$2D$ Hubbard model, no indication of ferromagnetism was also found
at $U=8t,\, \delta=0.35$ and $ T=0.3t$ on a $6\times 6$ square lattice in \cite{hirsch1}.
\end{thebibliography}
\end{document}